\documentstyle[11pt,aaspp4,psfig]{article}

\received{1997 January 17}
\accepted{1997 March 18}

\begin{document}
\title{Time-Resolved Ultraviolet Observations of the Globular Cluster X-ray 
Source in NGC\,6624: The Shortest Known Period Binary System
\footnote{Based on observations with the NASA/ESA Hubble Space Telescope 
obtained at the Space Telescope Science Institute, which is operated by the 
Association of Universities for Research in Astronomy, Inc., under NASA
contract NAS5-26555.}}

\author{Scott F. Anderson, Bruce Margon, and Eric W. Deutsch}
\affil{Astronomy Department, University of Washington, Box 351580, Seattle, 
WA  98195}
\affil{anderson, margon, and deutsch@astro.washington.edu}

\author{Ronald A. Downes}
\affil{Space Telescope Science Institute, 3700 San Martin Drive, Baltimore, 
MD  21218}
\affil{downes@stsci.edu}

\author{Richard G. Allen}
\affil{Steward Observatory, University of Arizona, Tucson, AZ 85721; 
rallen@as.arizona.edu}
\centerline{\it Accepted for publication in The Astrophysical Journal 
(Letters), vol. 482, 1997 June 10 issue}

\begin{abstract}
Using the Faint Object Spectrograph (FOS) aboard the {\it Hubble Space 
Telescope}, we have obtained the first time-resolved spectra of the King et al. 
ultraviolet-bright counterpart to the 11-minute binary X-ray source in the core 
of the globular cluster NGC\,6624. This object cannot be readily observed in 
the visible, even from {\it HST}, due to a much brighter star superposed 
$<0.1''$ distant. Our FOS data show a highly statistically significant UV flux
modulation with a period of 11.46$\pm$0.04 min, very similar to the 685~s 
period of the known X-ray modulation, definitively confirming the association 
between the King et al. UV counterpart and the intense X-ray source. The UV 
amplitude is very large compared with the observed X-ray oscillations: X-ray 
variations are generally reported as 2--3\% peak-to-peak, whereas our data 
show an amplitude of about 16\% in the 126--251 nm range. A model for the 
system by Arons \& King predicts periodic UV fluctuations in this 
shortest-known period binary system, due to the cyclically changing aspect of 
the X-ray heated face of the secondary star (perhaps a very low mass helium 
degenerate). However, prior to our observations, this predicted modulation has 
not been detected. Employing the Arons \& King formalism, which invokes a 
number of different physical assumptions, we infer a system orbital inclination 
$35^{\circ} \lesssim i \lesssim 50^{\circ}$. Amongst the three best-studied 
UV/optical counterparts to the intense globular cluster X-ray sources, two
are now thought to consist of exotic double-degenerate ultrashort period 
binary systems.

\end{abstract}

\keywords{X-rays: stars --- binaries: close --- globular clusters: general}

\section{Introduction}

In some respects, the X-ray source X1820-30 in NGC\,6624 is prototypical 
amongst the dozen highly luminous ($L_{x} \gtrsim 10^{36}$ erg s$^{-1}$) X-ray 
sources known in the Milky Way globular clusters. X1820-30 has been studied 
intensively in X-rays for several decades since its initial observation by 
{\it Uhuru} and SAS-3 (\markcite{Gia74}Giacconi et al. 1974; 
\markcite{JC79}Jernigan \& Clark 1979), was the first known X-ray burster 
(\markcite{Gri76}Grindlay et al. 1976), and shows quasi-periodic X-ray 
oscillations (\markcite{SWP87}Stella, White, \& Priedhorsky 1987). On the 
other hand, X1820-30 is truly exceptional in at least one X-ray characteristic: 
it shows a low amplitude ($\sim 3\%$ peak-to-peak), but highly coherent, X-ray 
modulation with a period of 11 min (\markcite{SPW87}Stella, Priedhorsky, \& 
White 1987; hereafter, SPW). The stability of this 11 min X-ray modulation is 
strongly indicative of an orbital period, and hence X1820-30 is thought to 
have the shortest known orbital period of {\it any} known binary system (SPW). 

The compact accreting binary components in the intense globular 
cluster X-ray sources are generally thought to be neutron stars: this is 
suggested by the presence of X-ray bursts and by estimates of 
$\sim$1.5 M$_{\odot}$ for total system masses based on the typical 
displacement of the X-ray sources from the globular cluster cores
(\markcite{Gri84}Grindlay et al. 1984). In the case of X1820-30 in NGC\,6624, 
with its suspected 11 min orbital period, such a binary system must necessarily 
be tiny, with a binary separation of only a few tenths of $R_{\odot}$:
separation $\sim 1.3\times 10^{10}[M_{ns}(1+q)]$ cm, where $M_{ns}$ is the 
the neutron star mass in units of 1.4~$M_{\odot}$, and $q$ is the mass ratio 
(\markcite{AK93}Arons and King 1993; hereafter, AK). Indeed the only 
mass-losing companion that will fit within such a tight binary system must be 
a degenerate dwarf (e.g., SPW; \markcite{V87}Verbunt 1987; AK) of very low
mass: a plausible companion is a Roche-lobe filling helium degenerate (white) 
dwarf with mass $\sim$ 0.07 M$_{\odot}$ and radius $\sim 0.03$R$_{\odot}$.
In such a system, gravitational radiation alone can drive sufficient mass
transfer to account for the high observed X-ray luminosity (SPW).

Although X-ray investigations alone have provided essential clues to the
nature of the luminous globular cluster sources, even the initial 
identification (much less follow-up) of the optical/UV counterparts has proved 
frustratingly difficult for ground-based investigations due to the extreme 
crowding in most globular cluster cores. A notable exception is the optically 
bright counterpart AC\,211 in M\,15, whose high optical emissivity permitted 
early identification and study from the ground (\markcite{ALT84}Auri\'ere, 
Le F\`evre, \& Terzan 1984; \markcite{CJN86}Charles, Jones, \& Naylor 1986).
However, {\it Hubble Space Telescope} ({\it HST}) has markedly improved this 
situation in several globular clusters, and one of the notable early successes
was the identification (using FOC images) of a highly ultraviolet object in 
the core of NGC\,6624 as the likely counterpart to X1820-30 
(\markcite{Kin93}King et al. 1993). A substantial majority of all the flux from 
the cluster at $\sim$1400 \AA\ is attributable to just this one star. 
Groundbased study of this system is nearly infeasible, as the UV object lies 
only 0.08$''$ from a presumably unrelated superposed star that is brighter than 
the King et al. counterpart in the optical.

AK provided a detailed model for the X1820-30 system. They invoke the 
previously suggested double-degenerate binary components, but their model also 
explains how the very intense emission at X-ray and UV wavelengths are likely 
intimately related. They argue that a Rayleigh-Jeans tail of emission from the 
X-ray emitting region of the system would be grossly inadequate to directly 
explain the high observed UV flux; the UV emission must therefore arise from 
reprocessing of X-rays intercepted by the companion star and the accretion 
disk, with the bulk of the UV emission arising in reprocessing from the disk. 
Moreover, although the King et al. FOC data did not permit a sensitive search 
for UV variability, AK nonetheless further {\it predicted} that the UV flux 
should exhibit a modest ($\sim$10\%) modulation at the 11 min X-ray period, as 
the X-ray heated face (heated to T$\sim 10^5$K) of the low-mass degenerate 
companion is alternately visible to and hidden from the observer in its orbit 
about the neutron star.

Here, we report on our ultraviolet {\it HST} Faint Object Spectrograph (FOS) 
observations of the King et al. counterpart in NGC\,6624. Although the 
time-averaged UV spectrum reveals no strong features, our FOS data are 
time-resolved and provide the first sensitive search for UV modulations at the 
11 min X-ray period (see also \markcite{And96}Anderson et al. 1996). The 
{\it HST} observations are described and the time-averaged UV spectrum 
presented in $\S$2. Our analysis of the FOS data for evidence of an 11 min UV 
modulation is described in $\S$3. Some implications of our results are 
discussed in $\S$4, especially with reference to the AK model.

\section{{\it HST} Observations and the Time-Averaged UV Spectrum}

On 1996 May 1 and 2 UT, we obtained low-resolution ultraviolet 
spectrophotometry from {\it HST} of the King et al. candidate. The data were 
taken with the G160L grating and FOS blue detector (broad 1150-2500~\AA\ 
coverage, but low spectral resolution $\lambda/\Delta\lambda=250$) with the 
0.4$''$ paired apertures. Eight-hundred separate FOS spectra were taken in 
``RAPID" mode, with a new spectrum taken every $\sim$15 s while the target was 
visible from {\it HST}; the data were collected during 21 distinct cycles of 
the 11 min X-ray period (and span 36 cycles). The FOS spectra have undergone 
the standard reductions performed on all such data processed through the 
STScI's pipeline reduction software.

The observed time-averaged UV spectrum is displayed in the upper panel of 
Fig. 1., and the lower panel shows the dereddened time-averaged spectrum 
assuming a reddening to the cluster of $E(B-V)=0.25$ (\markcite{RML93}Rich, 
Minniti, \& Liebert 1993). The observed flux agrees reasonably well with the 
earlier FOC estimate by \markcite{Kin93}King et al. (1993): the value at 
1400 \AA\ is only about 35\% lower in our FOS spectrum than that estimated by 
King et al. based on broadband FOC images (and of course the object may 
exhibit long-term variability in any case). The time-averaged observed spectrum 
is largely featureless at the low resolution and modest S/N of these FOS data, 
except for Ly$\alpha$ and the 2200 \AA\ interstellar feature. (There could also 
be extremely weak NV $\lambda\lambda$ 1238, 1242 in emission, but our current 
data lack adequate S/N and resolution to confidently confirm or rule this out.)
The Ly$\alpha$ emission is consistent with being largely geocoronal, as the 
total line strength is comparable to the geocoronal emission seen in the 
sky-aperture, but we cannot rule out some contribution from (or even net
Ly$\alpha$ absorption toward) the binary system. The adopted cluster reddening 
successfully removes the bulk of the prominent 2200~\AA\ interstellar feature, 
and the dereddened FOS spectrum of Fig. 1 (lower panel) is fit well by a 
power-law $f_{\lambda} \propto \lambda^{-3.0}$. The latter is in good agreement 
with the slope of $-3.2$ estimated by \markcite{Kin93}King et al. (1993) from 
their comparison of dereddened optical and UV broadband FOC images.

\section{Time-Resolved UV FOS Spectrophotometry}

Information on UV spectral variability at a variety of timescales, and in a 
variety of different UV wavelength bandpasses, may be derivable from these FOS 
data. However, especially motivated by the AK model prediction, our initial 
focus is on the most straightforward (and potentially best S/N) search for 
broadband UV variability at the 11 min X-ray period. We do this broadband 
search merely by summing together the UV counts collected in 180 diodes of the 
FOS blue detector that sample the 1st-order G160L spectrum from 1265~\AA\ to 
2510~\AA. This broadband UV count summation is done for each of the 800 
individual ``RAPID" readouts. Note that this wavelength bin purposefully starts 
conservatively longward of Ly$\alpha$, which is plausibly mainly of geocoronal 
origin. After correction for the detector noise background, we obtain in this 
way a simple measure of the broadband UV lightcurve of the system. The time 
resolution of these UV lightcurve data is $\sim$15 s, but the data are not 
uniformly sampled over the entire span of the {\it HST} observations due to 
earth occultations.

We use both Fourier power spectra (e.g., \markcite{RLD87}Roberts, Leh\'ar, \& 
Dreher 1987) and phase dispersion minimization (PDM) approaches 
(\markcite{S78}Stellingwerf 1978) to search for evidence of a broadband UV 
periodicity. For example, the upper panel of Fig. 2 shows a normalized Fourier 
power spectrum for the 1265-2510 \AA\ broadband lightcurve data. For both 
Fourier and PDM approaches the strongest period detected is centered near 
11.47 min. The UV broadband period is very similar to the known 685~s X-ray 
modulation (agreeing to within the expected Nyquist frequency resolution 
limitations for our observations, which span $\approx$400 min). It is clear 
that the statistical significance of our detection is very high indeed. The 
probability of an accidental peak due to random noise in a given bin of a 
Fourier power spectrum normalized to unity in the continuum is exponentially 
distributed, implying extremely small accidental occurrence probabilities
(formally of order $e^{-39}$) in the case of our Fig. 2.  Although this 
calculation cannot be taken literally, as the continuum normalization depends 
somewhat on the range of selected frequencies, our detection of the UV 
modulation is obviously firm. Hence, these FOS observations provide the first 
detection of an 11 min modulation from X1820-30 outside the X-ray regime. 

Additionally, we have least-squares fitted a simple sinusoid to the 800 data 
points of our broadband UV lightcurve. This nonlinear least-squares approach 
yields the following best-fit parameter values and associated {\it formal} 
errors, once again reaffirming the high confidence of the detection of an 
11 min modulation in these data: half-amplitude of 7.8$\pm0.7$\%, period of 
11.46$\pm0.02$ min, and epoch of maximum light at (heliocentric) 
$JD=2450205.1926 \pm0.0002$. The broadband UV lightcurve folded with these 
best-fit period/epoch values, and with lightcurve data-points further binned 
in phase to reduce counting noise, is shown in the lower panel of Fig. 2. While 
the folded and phase-binned UV lightcurve appears reasonably fit by a sinusoid,
a more complicated lightcurve is not ruled out (e.g., in fitting a sinusoid
to the full unbinned lightcurve data, the reduced chi-square is $\approx 1.4$).

The strength with which the 11 min UV oscillation is detected across the broad 
1265-2510 \AA\ band suggests the feasibility of detecting a 
wavelength-dependence to the modulation. In various trials, we subdivided the 
FOS broad UV spectral coverage into finer wavelength bins, generated separate 
lightcurves for each bin, and reran the Fourier, PDM, and non-linear least 
squares analyses separately for each lightcurve. For example, in one particular 
trial we subdivided the FOS spectral coverage into 4 wavelength bins 
(each about 300 \AA\ wide), yielding 4 distinct lightcurves derived from bins 
respectively centered near 1400~\AA, 1700 \AA, 2000 \AA, and 2300~\AA. In all 
trials, the lightcurve associated with each wavelength bin yielded a 
significant detection of the 11 min oscillation. But curiously, {\it within} 
each trial the values of the best-fit periods for two (or more) of the 
different-wavelength lightcurves were discrepant at the $\sim 2\sigma$ level 
from one another, taking the formal least-squares error bars at face value. 
This suggests that either the formal least-squares error bars on the period 
are artificially small, or that there is some real (but at best quite 
marginally detected) variation of the observed modulation period with 
wavelength. As the period and epoch in such a fit are strongly correlated, this 
might alternatively suggest a variation of the phase epoch with wavelength.
Therefore, we hereafter conservatively adopt the typical $rms$ deviations about 
the means seen in these various trials as more representative of the actual 
uncertainties for the period and epoch of the UV modulation: these 
deviations are $\pm$0.04 min for the period and $\pm$0.8 min 
($\pm 0.0006$~days) for the phase epoch. From these same trials, we also find 
that our data are compatible with, but do {\it not} require, a small trend in 
UV oscillation amplitude with wavelength. For example, for the aforementioned 
trial subdividing into 4 wavelength bins (each $\sim$300~\AA\ wide), the 
respective best-fit sinusoid half-amplitudes of the 4 lightcurves are 
8.8$\pm$1.5\%, 8.5$\pm$1.3\%, 7.7$\pm$1.4\%, and 6.9$\pm$1.0\%.

\section{Discussion and Conclusions}

The time-averaged spectrum of Fig. 1 is clearly that of a highly ultraviolet
object. This UV spectrum is consistent with an approximately power-law 
shape---nearly Rayleigh-Jeans in slope but somewhat more modest, as previously
suggested by King et al. based on their broadband FOC images---after correction 
for plausible reddening to the cluster. As they discuss, such a broad spectral 
energy distribution is in approximate agreement with the predictions of the 
AK model in which the bulk of the UV (time-averaged) flux arises from 
reprocessing of X-radiation off the accretion disk. The consistency of the 
required reddening correction needed to remove the 2200 \AA\ feature from our 
FOS spectrum clearly locates this UV object at a distance comparable to that 
of the globular cluster NGC\,6624 (confirming at least that the UV-object 
cannot be markedly foreground).

At the most basic level our detection of the 11 min UV periodicity---a 
periodicity previously known only in X-rays---confirms unequivocally that the 
King et al. UV-bright object is most certainly the counterpart to the luminous 
X-ray source in NGC\,6624. We hasten to point out that only the most cynical 
might have questioned this identification in any case, given the extraordinary 
ultraviolet time-averaged flux detected by King et al.

As noted in the introduction, the AK model for the X1820-30 system invokes but 
expands upon the previously suggested (e.g., SPW; \markcite{V87}Verbunt 1987) 
double-degenerate binary model (e.g., helium degenerate dwarf losing mass 
onto an accretion disk around a neutron star companion). Moreover, AK
{\it predicted} that the UV flux should exhibit a $\sim$5--10\% modulation at 
the 11 min X-ray period, as the X-ray heated face of the low-mass degenerate 
companion is alternately presented and hidden from the observer in its orbit 
about the neutron star. Our detection of an 11 min UV modulation with 
approximately the predicted amplitude must be viewed as strongly suggestive 
that at least some aspects of the AK double-degenerate model are correct. 

If we adopt the formalism of the AK model, the observed amplitude of the UV 
modulation provides information on the binary orbital inclination angle $i$. 
The modulation relation derived by AK predicts a half-amplitude of 
$\epsilon_{\nu}$sin(i)/[cos(i)+$\epsilon_{\nu}$sin(i)], where 
$\epsilon _{\nu}$ is the ratio of the low-mass degenerate dwarf's X-ray 
reprocessing luminosity to the reprocessing luminosity of one face of the 
accretion disk. An approximate value of $\epsilon_{\nu}$ may be obtained from 
Fig. 2 of AK; we estimate $\epsilon_{\nu} \sim 0.09$ at 1980 \AA, where the 
latter wavelength corresponds to the count-weighted mean wavelength of our 
1st-order FOS spectrum. The UV oscillation amplitude measured from our data 
thus corresponds to $i\approx 43^{\circ}$ ($-9^{\circ},+8^{\circ}$) in the AK 
model, where the quoted uncertainty considers only the {\it formal} 3$\sigma$ 
limits on the best-fit sinusoid amplitude. (These formal errors on the 
inclination, of course, do not account for systematic uncertainties associated 
with simplifying model assumptions, e.g., shadowing of the secondary by the 
disk is neglected).

The AK formalism also predicts an increasing oscillation amplitude toward 
shorter wavelengths, with $\epsilon_{\nu}$ and the oscillation amplitude 
$\sim1.4\times$ larger at 1400~\AA\ than at 2300 \AA. Our data (see \S 3) yield 
a best-fit amplitude near 1400 \AA\ which is $\approx 1.3\pm0.3$ times larger 
than that near 2300 \AA, and thus are also consistent with this aspect of the 
AK model, but lack sufficient precision to provide a definitive test.

In closing, it is also of interest to note that based on WFPC2 photometric 
monitoring in the near-UV, \markcite{Hom96}Homer et al. (1996) have recently 
concluded that a second globular cluster X-ray source counterpart, ``Star S" 
in NGC\,6712 (\markcite{Bai91}Bailyn et al. 1991; \markcite{AKM92}Auri\'ere \& 
Koch-Miramond 1992; \markcite{And93}Anderson et al. 1993), is also a 
double-degenerate with an ultrashort 20.6 min binary orbital period. Moreover, 
the (albeit very low S/N) dereddened FOS ultraviolet spectrum of ``Star S" in 
NGC\,6712 displayed in \markcite{DAM96}Downes, Anderson, \& Margon (1996) also 
rises steeply toward shorter wavelengths, and hence the ``Star~S" UV spectral 
energy distribution may at least resemble that of the King et al. counterpart 
in NGC\,6224. At the very least it is a curious result---and perhaps a much 
more fundamental one---that among the 3 best-studied counterparts (AC\,211 in 
M\,15, ``Star S" in NGC\,6712, and the King et al. object in NGC\,6624) to the 
intense globular cluster X-ray sources, two of the objects are now thought to 
be unusual double-degenerate systems with ultrashort binary periods. Such 
systems are evidently far more rare among the intense binary X-ray sources in 
the field. Further identifications and follow-up of the counterparts to the 
intense X-ray sources in other globular clusters (e.g., 
\markcite{Deu96a}\markcite{Deu96b}Deutsch et al. 1996a, 1996b) may reveal 
whether such double-degenerates are truly overabundant, or merely that this 
initial hint is an artifact of the extremely small number of counterparts 
closely studied thus far.

\acknowledgments
We thank J. Arons, B. Beck-Winchatz, D. Hoard, and A. Silber for comments
and assistance. This work has been supported by NASA Grant NAG5-1630.

\newpage

\begin{figure}
\hskip 1.0in
\psfig{figure=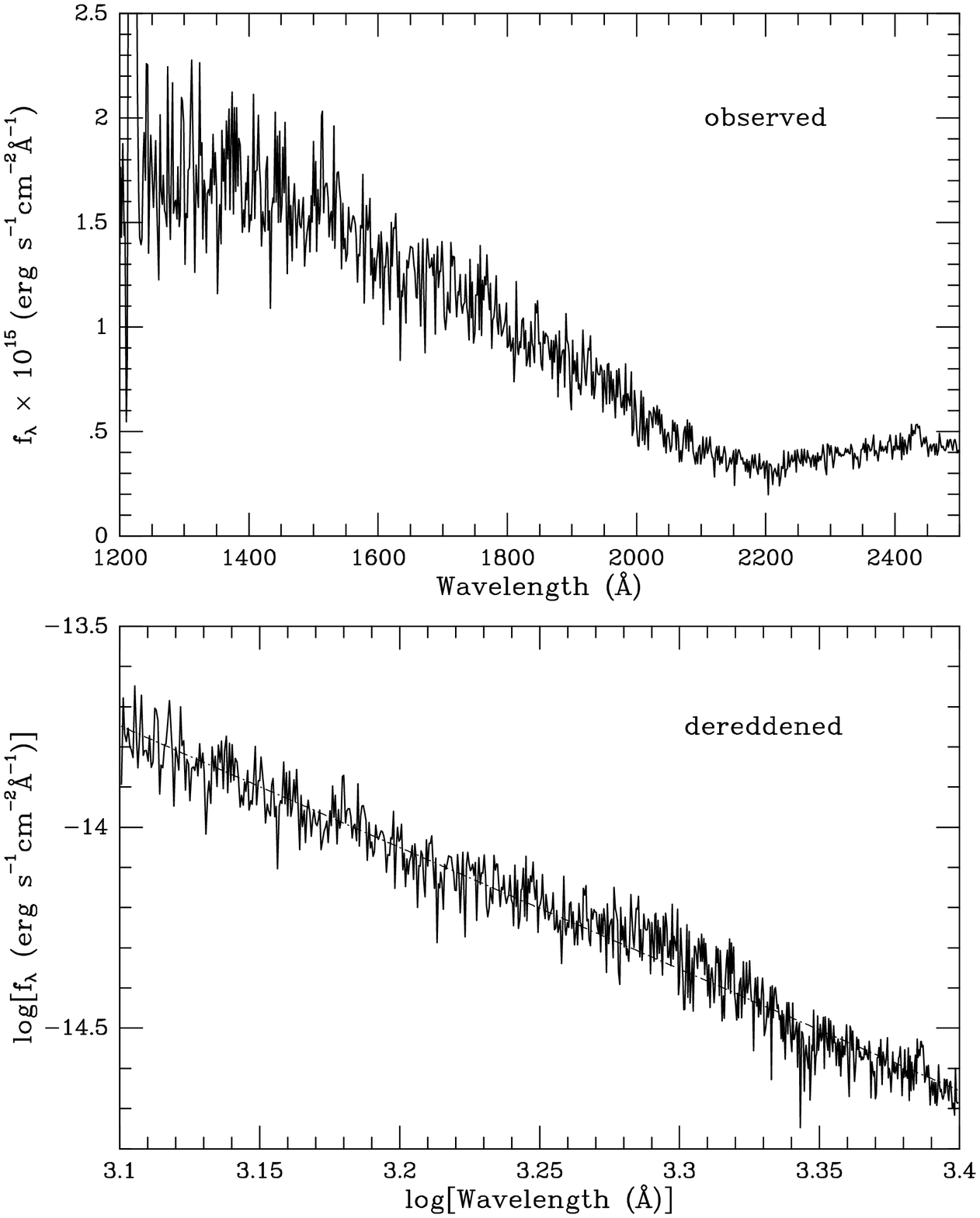,height=18truecm,width=12truecm,angle=0}
\caption[]{The time-averaged UV FOS spectrum of the King et al. counterpart 
to the X-ray source X1820-30 in the globular cluster NGC\,6624. {\it Upper
panel}: the observed spectrum is clearly that of a highly ultraviolet object, 
but few (if any) strong spectral features intrinsic to the object are seen. 
{\it Lower panel}: the dereddened spectrum, assuming $E(B-V)=0.25$ as 
appropriate for NGC\,6624; this spectrum is fit well by a power-law
$f_{\lambda} \propto \lambda^{-3.0}$ (dashed-line), indicating a strongly 
rising spectrum slightly less dramatic than Rayleigh-Jeans.}
\end{figure}

\begin{figure}
\hskip 1.0truein
\psfig{figure=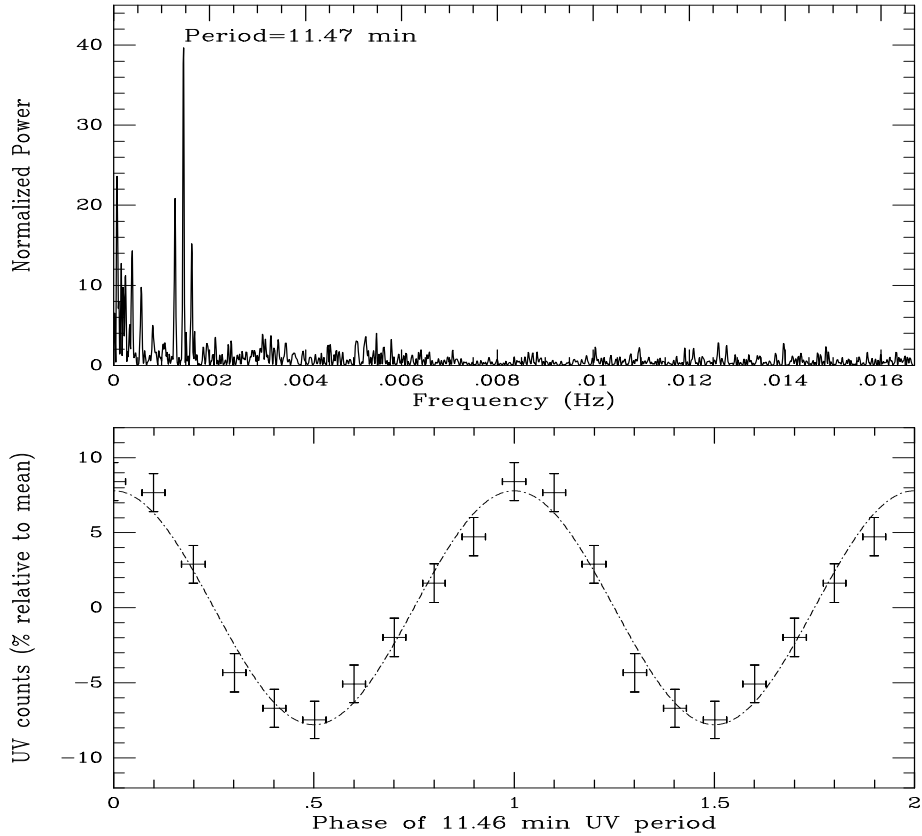,height=12truecm,width=12truecm,angle=0}
\caption[]{Fourier and least-squares analyses of the broadband FOS UV 
lightcurve. Eight hundred separate spectra were taken with time resolution 
$\sim$15 s. A simple measure of the UV broadband lightcurve was obtained by 
summing the counts within each of the 800 spectra across detector diodes 
corresponding to 1265-2510~\AA. Fourier, PDM, and least-squares approaches all 
yield very strong detections of an 11 min periodicity in the UV lightcurve. 
{\it Upper panel}: the normalized Fourier power spectrum of the UV lightcurve.
The strongest feature, of very high significance, is centered at 11.47 min; 
this UV period is very similar to that of the known X-ray period, confirming 
unequivocally that the X-ray source X1830-20 and the King et al. UV-bright 
object are associated. (The strong sidebands are beats with the 96 min 
{\it HST} orbital period). {\it Lower~panel}: the UV lightcurve folded with a 
best-fit period and phase epoch (phase 0 chosen to be UV max) estimated from a 
non-linear least-squares fit. Ten independent phase bins across the 11 min UV 
period are displayed (but two cycles are shown with redundant data for 
clarity). {\it Horizontal error bars}: standard deviation in phase sampled by 
each of the 10 phase bins; {\it vertical error bars}: $\sqrt{N}$ expectations 
for the count uncertainties; {\it dashed curve}: best-fit sinusoid, with 
parameters obtained from fitting to all 800 data points of the UV lightcurve. 
This simple model fit yields a very large amplitude of 16\% peak-to-peak in 
the UV (compared to $\sim3\%$ for X-rays).}

\end{figure}
\clearpage
\end{document}